\documentclass[11pt]{article}

\usepackage{appendix}
\usepackage{graphicx}
\usepackage{lscape}
\usepackage{verbatim}
\usepackage{color}
\usepackage{epstopdf}
\usepackage{amsmath}
\usepackage{amsfonts}
\usepackage{latexsym}
\usepackage{booktabs}
\usepackage{algorithm}
\usepackage{algorithmic}
\usepackage{amsmath}
\usepackage{txfonts}
\usepackage{subfigure}
\usepackage{amssymb}
\usepackage{subfigure}
\usepackage{multicol}
\usepackage{multirow}
\usepackage{colortbl}
\usepackage{slashbox}
\usepackage{hhline}
\usepackage{setspace}
\usepackage[hmargin=1in,vmargin=1in]{geometry}

\begin{document}

\noindent \textbf{Flexible multivariate marginal models for analyzing multivariate longitudinal data, with applications in R} \\

\noindent \"{O}zg\"{u}r Asa$\mbox{r}^{ \ a, *}$ \let\thefootnote\relax\footnotetext{$^*$ Corresponding author. Tel.: +44 (0) 1524 593519. E-mail address: o.asar@lancaster.ac.uk (\"{O}. Asar)}, \"{O}zlem \.{I}l$\mbox{k}^{ \ b}$ \\

\noindent $^a$ CHICAS, Lancaster Medical School, Faculty of Health and Medicine, Lancaster, LA1 4YG, UK.\\
\noindent $^b$ Department of Statistics, Faculty of Arts and Sciences, Middle East Technical University, Ankara, 06800, Turkey.\\

\noindent \textbf{ABSTRACT}\\

\noindent Most of the available multivariate statistical models dictate on fitting different parameters for the covariate effects on each multiple responses. This might be unnecessary and inefficient for some cases. In this article, we propose a modeling framework for multivariate marginal models to analyze multivariate longitudinal data which provides flexible model building strategies. We show that the model handles several response families such as binomial, count and continuous. We illustrate the model on the Mother's Stress and Children's Morbidity data set. A simulation study is conducted to examine the parameter estimates. An R package {\bf mmm2} is proposed to fit the model. \\


\noindent \textbf{1. \ \ \ \ \ \ Introduction} \\

Longitudinal data include observations which are collected repeatedly over time from same subjects, and this type of data is common in many research areas, e.g., medical studies ~\cite{yates13, jang13}. Often, multiple response variables are collected on each subjects. These repeated measurements are typically dependent and three dependence structure can be mentioned: within-response, between response and cross-response temporal dependencies. While the former corresponds to the dependence within each response variables across time, the latter ones corresponds to dependence between multiple responses at a specific time point and dependence between multiple responses at different time points, respectively.

Mother's Stress and Children's Morbidity (MSCM ~\cite{alexander86}) data set was collected with the aim of investigating the effect of mother's employment status on the pediatric care usage. The study included 167 mothers and their preschool children (aged between 18 months and 5 years) with no chronic disease. In MSCM data set, two binary response variables, specifically mother's stress status (0=absence, 1=presence) and her child's illness status (0=absence, 1=presence) were collected over 28 days, i.e., this is a bivariate longitudinal binary data set. Here, the dependence between mother's stress status (or child's illness status) at day $(t-m)$ and at day $t$, where $1 \leq m \leq t-1$, corresponds to within-subject dependence. Whereas the dependence between mother's stress and her child's illness at day $t$ corresponds to multivariate response dependence, the one between mother's stress at time $(t-m)$ and her child's illness at $t$, where $1 \leq m \leq t-1$, corresponds to cross-response temporal dependence.

Marginal models are the extensions of generalized linear models ~\cite{mccullagh89} to correlated data. They permit regression parameter interpretations that are free of the dependence structures and are useful when the interest is on population rather than individuals. In these models, dependence structures of longitudinal responses are of secondary interest. Nonetheless, the dependencies should be taken into account to draw valid statistical inferences. Since the specification of the multivariate distribution of the longitudinal responses is very complex and difficult even for the univariate response case, semi-parametric approaches for parameter estimation would be beneficial. ~\cite{liang86} proposed generalized estimating equations (GEE) for parameter estimation in univariate marginal models. With this approach, one only needs to specify the functional relationship between the mean response and the covariates, and the mean and variance of the responses. ~\cite{fitzmaurice04} reported that this semi-parametric approach is able to compete with the full likelihood-based ones in many cases. Two outstanding features of GEE are: 1) it is not restricted to any specific response family but handles several of them, 2) it yields consistent parameter and variance estimates even under misspecification of the dependence structure. The works of ~\cite{diggle02, molenberghs05, fitzmaurice09, ziegler11} include great literature for univariate marginal models and GEE. 

Analysis of multivariate longitudinal data has quite limited literature, especially for marginal models. Review of such methods with a general aspect, i.e., not only focused on marginal models, could be found in ~\cite{bandyopadhyay11, verbeke12}; the latter has a broad perspective while reviewing the existing methods. ~\cite{chaganty02} considered quasi-least square methods for multivariate models. ~\cite{shelton04} proposed models for binary data with GEE. ~\cite{asar13} generalized the work of ~\cite{shelton04} for other response families rather than binomial and proposed the R ~\cite{r12} package mmm ~\cite{mmm}. ~\cite{jia09} considered likelihood based models with common predictor effects for continuous data by using model selection tools. ~\cite{genolini13} considered joint clustering of multivariate individual trajectories. 

Traditional multivariate model formulation approach postulates the assumption of separate relationships between multiple responses and covariates, i.e., fitting separate regression coefficients for each of the relationships. However, in some cases all or some of these relationships might be very similar. This yields estimation of redundant regression coefficients and causes losses in the efficiencies of the parameter estimates. For instance, in the MSCM study, we have $2$ responses and $11$ covariates and the responses have similar relationships with $9$ of the covariates. The aforementioned traditional formulation would necessitate estimating $24$ separate marginal regression parameters (including intercepts) to completely specify the relationships between response means and covariates. However, inclusion of $9$ separate parameters for the similar relationships seems to be redundant. Therefore, we need flexible model building methodologies. ~\cite{ilk07} proposed an approach in multivariate models which assumes that all the covariate effects (including intercepts) are shared across multiple responses. Besides, their methodology permits fitting separate intercepts and covariate effects by including response type indicator variables and related interactions with the covariates in the design matrix. 

In this study, we propose a multivariate marginal model which extends the works of ~\cite{shelton04, asar13, ilk07} and shares the notion of ~\cite{jia09}. Its novel feature is that it permits fitting common as well as separate regression parameters for different responses within the multivariate marginal modeling framework. Parameterizing shared relationships are decided based on statistical significance of the interactions regarding response type indicator variables. Our model is valid for several response families such as binomial, Gaussian and Poisson. The parameters are estimated via GEE. We propose an R  package {\bf mmm2} ~\cite{mmm2} to fit the model.

The article is organized as follows. In Section 2, we introduce the proposed model and illustrate the related parameter estimation procedure. In Section 3, we apply the model on the MSCM data and give some illustrative R code snippets for implementation. In Section 4, we conduct a simulation study to illustrate the gains in the efficiencies. We close the article by discussion and conclusion placed in Section 5. \\

\noindent \textbf{2. \ \ \ \ \ \ Flexible multivariate marginal models} \\

\noindent \textit{\textbf{2.1 \ \ \ \ \ \ Model}} \\

The formulation of the model is given by

\begin{equation}
\label{eq:multimarg2}
g(E(Y_{itj}|X_{itj})) = g(\mu_{itj}) = X_{itj} \beta.
\end{equation}

Here, $Y_{itj}$ is the $j$th $(j=1, \ldots , k)$ response of subject $i$ $(i=1, \ldots , N)$ at time $t$ $(t=1, \ldots , n_i)$. $X_{itj}$ is the associated set of covariates which might be changing with time (time varying) or not changing with time (time invariant). $\beta$ is the vector of regression coefficients to be estimated and assumed to be shared across multiple responses. Put another way, we assumed that all the covariate effects as well the intercepts are shared across multiple responses. Nevertheless, we might allow multiple responses having their own intercepts by including the response types as indicator variables in the design matrix. Additionally, we might allow multiple responses having their own slopes by including the interactions between the response types and the covariates in the design matrix. These aspects will be clear when illustrated by examples in Section 3. Since inclusion of the response type indicator variables introduce covariates that depend on the response variable, we use the response index, $j$, while denoting the design matrix, $X_{itj}$. $g( . )$ is a known link function which linearizes the relationship between the covariates and the mean response. Possible choices include identity for continuous data, logit and probit for binary data and natural logarithm for count data. 

The association parameters are not explicitly specified in Equation \ref{eq:multimarg2}, since they are not of primary statistical interest, i.e., nuisance parameters. These parameters, in fact, are specified within the GEE approach while estimating the parameters.

The traditional model of ~\cite{shelton04, asar13} differs from our model by assuming common set of covariates and different set of regression coefficients for multiple responses, i.e., $g(E(Y_{itj}|X_{it}))=g(\mu_{itj})=X_{it} \beta_j$. In other words, this corresponds to $\beta_j \not \equiv \beta_{j^\prime}$ for $j \neq j^\prime$. However, the model which is the subject of this paper permits having $\beta_j \equiv \beta_{j^\prime}$ or $\beta_j \not \equiv \beta_{j^\prime}$. Differences between these models and the advantages of the current one will be discussed in Section 3 with applications. \\

\noindent \textit{\textbf{2.2 \ \ \ \ \ \ Parameter estimation}} \\

The parameter estimation process includes two main steps: 1) reconstruction of response and design matrices, 2) utilizing the original proposal of GEE ~\cite{liang86} afterwards. These steps are depicted below.

For the ease of understanding, let's assume a hypothetical longitudinal data set in which $k$ multivariate longitudinal responses and $p$ covariates are available. Also, let's assume that the longitudinal data is collected on $N$ subjects with $n_i$ repeated measurements, which would yield $M=\sum_{i=1}^{N} n_i$ total number of observations. The multivariate responses form an $M \times k$ matrix which can be denoted by $Y=(Y_1, \ldots, Y_N)^T$, where $Y_i=(Y_{i.1}, \ldots, Y_{i.k})^T$ and $Y_{i.j}=(Y_{i1j}, \ldots, Y_{in_ij})^T$. Similarly, the design matrix has a dimension of $M \times (p+1)$ and can be denoted by $X=(X_i, \ldots, X_N)^T$ where $X_{i}=(1, X_{i.1},\ldots, X_{i.p})^T$ with $1=(1, \ldots, 1)^T$ (having $n_i$ rows) and $X_{i.l}=(X_{i1l}, \ldots, X_{in_il})^{T}$. 

The multivariate response matrix is manipulated to construct a new one, $Y_{new}$, with a dimension of $(M \times k)  \times 1$, where $Y_{new}=(Y_{1,new}, \ldots, Y_{N,new})^{T}$ with $Y_{i,new}=(Y_{i1.}, \ldots, Y_{in_i.})$ and $Y_{it.}=(Y_{it1}, \ldots, Y_{itk})$. On the other hand, the reconstruction of the design matrix largely depends on the scientific interest. In other words, the manipulation depends on whether only the intercepts or both the intercepts and a certain subset or the whole set of covariate effects are to be separated across the responses. First, let's start with the case in which all the intercepts and covariate effects are to be shared across responses. This indicates extending $X$ to $X_{new}$ which has a dimension of $(M \times k) \times (p+1)$, where $X_{new}=(X_{1,new} , \ldots , X_{N,new})^{T}$ with $X_{i,new}=(\mbox{replicate}(X_{i1.})^T, \ldots, \mbox{replicate}(X_{in_i.})^T)$. The replicate function creates $k \times (p+1)$ matrices for which the rows are identical and equal to $X_{it.}$ where $X_{it.}=(1, X_{it1}, \ldots, X_{itp})$. If the intercepts or both the intercepts and a certain subset or the whole set of covariate effects are to be separated across the responses, we need to create additional covariates and append them to $X_{new}$. To separate the intercepts for multiple responses, we need to create $(k-1)$ response type indicator variables. Furthermore, to separate the covariate effects (a certain subset or all of them), we need to create $(k-1) \times p^*$ covariates for the interactions of the indicator variables and the covariates. These additional covariates are inserted to the right hand side of $X_{new}$, i.e., eventually, we have a design matrix of size $(M \times k) \times (p+1+(k-1)+(k-1) \times p^*)$, where $p^*$ is the number of covariates for which the effects on multiple responses are to be separated and $p^* \leq p$. Here, we shall note that if $p^* = p$, the model corresponds to the traditional model mentioned above. 

The regression parameter estimates, $\hat{\beta}$, are obtained via the GEE approach by considering $Y_{new}$ as a univariate response matrix and $X_{new}$ as the associated design matrix. We do not give the details of the application of GEE methodology to multivariate marginal models here, since ~\cite{asar13} illustrated a very similar application in Section 2.2 of their paper. 

In GEE approach, one does not need to fully specify the distribution of the responses. However, only the functional relationship between the response mean and response variance, and the one between the response mean and the covariates need to be correctly specified. It is well known that GEE yields consistent regression parameter estimates and their variances even when one selects an incorrect working correlation structure. On the other hand, correct choices of such structures would increase the efficiency of the estimates. Since our parameter estimation methodology directly adapts the GEE approach, these features are inherited to our estimates. We will discuss them in Section 3 with applications.

The estimation procedure introduced in this section is implemented via the R function \verb mmm2  which is available under the R package {\bf mmm2}. The \verb gee  function under the {\bf gee} package ~\cite{carey12} is utilized within \verb mmm2 . Robust and model based estimates of the standard errors and Z statistics are produced at the same time. We skip the details of the arguments of \verb mmm2  here, since these details are readily provided in the user manual of {\bf mmm2}. Nevertheless, we provide some sample R code snippets in Section 3 together with the modeling formulations and results. \\

\noindent \textbf{3. \ \ \ \ \ \ Mother's Stress and Children's Morbidity Study applications} \\

\noindent \textit{\textbf{3.1 \ \ \ \ \ \ Data}} \\

In Mother's Stress and Children's Morbidity Study (MSCM), 167 mothers and their preschool children were enrolled for 28 days. At baseline all the demographic variables regarding the mother and her children were collected by an interview. Then, the mothers were required to report their stress status (0=absence, 1=presence) and their children's illness status (0=absence, 1=presence) by telephone. All the variables can be found in Table 1. \\

\noindent [Insert Table 1 here.]\\

We considered only the period of day 17 to day 28 (a period of $12$ days), since the empirical investigation of the correlation structures suggested a weak correlation structure for the period of days 1 to 16 for both of the responses. To accommodate the specific features of the mothers and children in our models, we added the average response values of the neglected time period (days 1 to 16) as new covariates; see ``baseline stress" and ``baseline illness" in Table 1. Moreover, to measure the time effect in the mean stress and illness status, we added the standardized time information to our covariate list; see ``week" in Table 1. This version of the MSCM data set is available under the R package {\bf mmm2} with the name of \verb mscm . Analyses of the MSCM data set in the univariate marginal modeling framework could be found in the works of ~\cite{diggle02, zeger86}. \\

\noindent \textit{\textbf{3.2 \ \ \ \ \ \ Results}} \\

While building parsimonious models, the first step might be building the most general model which assumes that all the intercepts and all the covariate effects on multiple responses are different. Following the results of this model, one can decide which covariate effects and whether the intercepts are to be shared across the multiple responses. The most general model for the MSCM data via the use of {\em logit} link function could be given by

\begin{align}
\label{eq:most.gen.mod1}
\mbox{logit}(P(Y_{itj}=1|X_{itj})) = \beta_{0}+\beta_{1}*married_{i}+\cdots+\beta_{11}*week_{t}+ \beta_{12}*rtype_{j}+ \nonumber \\
\beta_{13}*(married_{i}*rtype_{j})+\cdots+\beta_{23}*(week_{t}*rtype_{j}).
\end{align}

Let's assume rtype=0 for response=stress and rtype=1 for response=illness. Then, the model given in Equation \ref{eq:most.gen.mod1} indicates the following models:

\begin{align}
\label{eq:most.gen.mod2}
\mbox{logit}(P(Y_{it1}=1|X_{it1}))=\beta_{0}+\beta_{1}*married_{i}+\cdots+\beta_{11}*week_{t}
\end{align} 

\noindent and

\begin{align}
\label{eq:most.gen.mod3}
\mbox{logit}(P(Y_{it2}=1|X_{it2}))=(\beta_{0}+\beta_{12})+(\beta_{1}+\beta_{13})*married_{i}+ \cdots + (\beta_{11}+ \beta_{23})*week_{t}
\end{align} 

\noindent for stress and illness, respectively. As it can be seen from Equations \ref{eq:most.gen.mod2} and \ref{eq:most.gen.mod3}, the intercept and covariate effects on mother's stress and child's illness variables are different. For instance, while the effect of mother's marriage status (married) on mother's stress is assumed to have a magnitude of $\beta_1$, it is assumed to have a magnitude of $(\beta_{1}+\beta_{13})$ on child's illness.
 
Related results under exchangeable working correlation structure are displayed in Table 2 under the Model 1 column. We can decide whether multiple responses to have their own parameters by investigating the significance of the interaction terms. In other words, failing to reject the null hypothesis in the hypothesis test of $H_0:\beta_{s}=0 \ vs \ H_0:\beta_{s} \neq 0$, where $\beta_s$ corresponds to an interaction coefficient of \emph{rtype} with a covariate, i.e., $s=13, \ldots, 23$ in Equation \ref{eq:most.gen.mod1}, would direct us on deciding whether the related covariate to have shared effect on multiple responses. Similarly, we can decide whether the responses to have a shared intercept by testing the significance of the coefficient of the \emph{rtype}, i.e., testing $H_0:\beta_{12}=0 \ vs \ H_1:\beta_{12} \neq 0$ in Equation \ref{eq:most.gen.mod1}. Results of Model 1 showed that we can allow multiple responses to have shared slopes for all the covariates with 95\% confidence, except for the size of the household (housize) and baseline stress (bstress). For instance, this might be read as, a mother's stress and her child's illness evolved similar to each other across time (week), the odds ratios for these responses were $0.65 \ (=exp(-0.43))$ and $0.83 \ (=exp(-0.43+0.24))$ with respect to successive days, respectively. We can build a more parsimonious model, a model with 9 less parameters, by omitting the aforementioned insignificant interactions. This model is given by

\begin{align}
\label{eq:pars.model1}
\mbox{logit}(P(Y_{itj}=1|X_{itj})) = \beta_{0}+\beta_{1}*married_{i}+\cdots+\beta_{11}*week_{t} + \beta_{12}*rtype_{j}+ \beta_{13}*(housize_{i}*rtype_{j})+ \nonumber \\ 
\beta_{14}*(bstress_{t}*rtype_{j}).
\end{align} 
 
We displayed related results under exchangeable working correlation structure in Table 2 under the Model 2 column. They indicated gains in the efficiencies for all the parameters compared to Model 1, since we estimated $9$ less parameters. For instance, standard error of response type indicator variable decreased to $0.31$ in Model 2 while it was $0.54$ in Model 1. We also reported the results of the model given in Equation \ref{eq:pars.model1} under unstructured working correlation structure in Table 2 under the Model 3 column. Comparison of the results of Model 2 and Model 3 reveals that the parameter estimates are consistent under different working correlation choices. We shall note that the model fitting algorithm, specifically Fisher-Scoring algorithm, did not converge to a solution for the most general model given in Equation \ref{eq:most.gen.mod1} under the unstructured working correlation matrix, since the model requires the estimation of $2*12 \choose 2$ = 276 different correlation parameters for the MSCM data set ~\cite{asar13}. Building a more parsimonious model, the one given in Equation \ref{eq:pars.model1}, enabled convergence to a solution under this working correlation matrix with estimation of same number of correlation parameters. \\
 
\noindent [Insert Table 2 here.]\\ 
 
The results of Models 1-3 could be easily translated to the ones for different responses. For instance, the regression parameter estimates could be obtained by plugging in the values of {\em rtype}, i.e., $(\hat{\beta}_{s}+ \hat{\beta}_{s^{\prime}} * rtype_j)$. Moreover, standard error estimates could be calculated by using the following well-known formula:  $\sqrt{var(\hat{\beta}_{s})+var(\hat{\beta}_{s^{\prime}})+2*cov(\hat{\beta}_{s},\hat{\beta}_{s^{\prime}})}$. Then, the calculation of $Z$ statistics are straightforward as the usual Wald type calculation. All of these results are presented in Table 1 of the web appendix of this paper available at http://www.lancs.ac.uk/pg/asar/web-appendix-mmmflex. For instance, under Model 2, while the estimate of the intercept for response=stress was $-2.33=-2.33+0*0.89$, it was $-1.34=-2.33+0.89$ for response=illness. Related standard error estimates were $0.36=\sqrt{0.13+0+2*0}$ and $0.41=\sqrt{0.13+0.09+2*(-0.03)}$, respectively. In Table 1 of the web appendix, we also included the percentage gains in the efficiencies, calculated as the percentage decreases in the robust standard error estimates in terms of Model 2 and Model 3 in comparison with Model 1. Results showed that there were considerable amount of gains. For instance, there was almost 33\% gain in the efficiency for the employment status of mothers (employed) for response=illness under Model 2. The gains of Model 3 seem to be slightly better compared to the gains of Model 2.\\

\noindent \textit{\textbf{3.3 \ \ \ \ \ \ Implementation}} \\
 
Model 1 can be fitted in R by the following script

\begin{verbatim}
# installing the package from CRAN
R> install.packages("mmm2")
# loading the package into R
R> library("mmm2")
# loading the MSCM data set
R> data(mscm)
# fitting Model 1
R> fit <- mmm2(formula=cbind(stress, illness) ~ married + 
+            education + employed + chlth + mhlth + race + 
+            csex + housize + bstress + billness + week,
+            data=mscm, id=mscm$id, rtype=TRUE, 
+            interaction=1:11, family=binomial, 
+            corstr="exchangeable")
\end{verbatim}

Here, while \verb rtype=TRUE  corresponds to the inclusion of response type indicator variable, \verb interaction=1:11  corresponds to the inclusion of interactions of the response type indicator variable with the first 11 covariates (all of the covariates) as new covariates. Model 2 can be fitted by a similar script with a little change of \verb interaction=c(8,9). Moreover, Model 3 could be built with an additional change of \verb corstr="unstructured"  for Model 2. The results could be displayed by the \verb summary()  function, e.g., the output can be displayed by \verb summary(fit)  for Model 1. 

We calculated the computational times required to fit these three models. While Model 1 took 0.34 seconds, Models 2 and 3 took 0.17 and 0.67 seconds, respectively on a personal computer with 4.00 GB RAM and 2.53 GHz processor. The reason why Model 3 took the longest time is that it was fitted under the unstructured working correlation matrix.

We considered application of the model presented in this paper on multivariate longitudinal count and continuous data sets, but we preferred not to include them here due to page limits. These data sets are available under the {\bf mmm2} package with names of \verb mlcd  and \verb  mlgd  for these response types, respectively. Moreover, the related R scripts to fit the models are available under the package manual. The model building strategies would be similar to ones illustrated for the MSCM data set. The only difference would be on the model formulations, i.e., {\it logit} should be replaced with {\it log} and {\it identity} link functions for count and continuous responses, respectively.

We fitted the traditional model of ~\cite{shelton04, asar13} and compared the results with Model 1 by using the {\bf mmm} package. We obtained identical results as expected. \\

\noindent \textbf{4. \ \ \ \ \ \ A simulation study} \\

\noindent \textit{\textbf{4.1 \ \ \ \ \ \ Data generation}} \\

We conducted a simulation study to investigate the bias and efficiency of the estimates of the proposed model. We reported mean, bias and mean squared error (MSE) of the estimates for this purpose.

Data were generated under the multivariate model of ~\cite{ilk07} with a probit link ~\cite{asar12}, to create within and between response dependencies while generating data from the marginal modeling framework. This model was proposed for multivariate longitudinal binary data. It is a marginalized multilevel model with three levels. The first level is nothing but a multivariate marginal model. The second and third levels are designed to capture the serial and multivariate response dependencies, respectively. Due to page limits, we do not give more details here; interested reader may refer to the cited references.

We mainly assumed that there are 300 subjects $(i=1, \ldots, 300)$ who were followed repeatedly over 3 time points $(t=1,2,3)$. We further assumed that two binary responses $(j=1,2)$ and two covariates were measured for each subject at each time point. In the following discussion, subscripts are suppressed whenever it is possible.  The relationship between the responses and the covariates were specified by

\begin{align}
\label{eq:sim.model}
P(Y=1|X)=\Phi( \beta_0 + \beta_1 *X_1+\beta_2* X_2+\beta_3* X_1*X_2+\beta_4* rtype+ \nonumber \\ 
\beta_5* X_1*rtype+ \beta_6* X_2*rtype+\beta_7* X_1*X_2*rtype),
\end{align} 

\noindent via the first level of the data generation model. Here, $\Phi$ denotes {\emph probit} link function which is defined as the cumulative distribution function of the standard normal distribution. $X_1$ was taken as a time-varying covariate which was generated by $X_{t,1}=\gamma_{0}+\gamma_{1}*X_{t-1,1}+\varepsilon_t$ for $t=2,3$ with $(\gamma_0,\gamma_1)=(0.2,0.5)$, $\varepsilon_1 \sim N(0,0.25^2)$, $\varepsilon_2 \sim N(0,0.15^2)$ and $X_{1,1} \sim N(0,0.4^2)$. $X_2$ was a time independent binary variable following a Bernoulli distribution with success probability of 0.5. The $rtype$ took 1 for the first response $(j=1)$ and 0 for the second response $(j=2)$. The regression parameters were selected as $(\beta_1,\beta_2,\beta_3,\beta_4,\beta_5,\beta_6,\beta_7)=(-0.5,0.5,0.9,0.6,0,0,0,0)$. This configuration corresponds to the case where intercept and all the covariate effects are shared across multiple responses. The association parameters of the data generation model, i.e., the parameters of the second and third level of the model, were set to have moderate within-response and between-response correlations. For instance, the means of the correlations were around 0.5 and 0.25 for these dependencies, respectively. Data sets were analyzed by two different models, the most parsimonious (true model for this simulation study) and the most general model, under several working correlation matrix choices. The simulation study was replicated 10,000 times. Probit analysis was achieved via \verb mmm2  function by setting the \verb family  argument to \verb family=binomial(link=probit) .\\ 

\noindent \textit{\textbf{4.2 \ \ \ \ \ \ Results}} \\

Results of the simulation study are represented in Table 3. We reported the results of two main models. The first one is the model which assumes that the intercepts and the covariate effects are shared across the bivariate responses. The results of this model were placed under the columns named Parsimonious. Note that this model is the true model in the sense of data generation. The second one assumes that the intercepts and covariate effects are all response specific. The result of this model were placed under the columns named Common. Both of these models were built under four different working correlation assumptions: unstructured, exchangeable, AR(1) and independence. All of the models yielded essentially unbiased regression parameter estimates under all of the working correlation assumptions. There seemed no apparent difference between the estimates of the two models under any working correlation matrix choices in terms of bias. However, in terms of MSE's the parsimonious model seemed to be outperforming the common model for all of the parameters. In fact, the MSE's were almost doubled. For instance, for $\beta_3$, while the former yielded estimates with an MSE of 0.051 under exchangeable structure, the latter model yielded estimates with an MSE of 0.096. However, the working correlation matrix choices seemed not to differ in terms of MSE's. For instance, for the same parameter, the MSE's of the former model were found to be 0.046, 0.053 and 0.054 for the unstructured, AR(1) and independence working correlation matrices. Interestingly, the common model yielded highest MSE's for $\beta_7$ when compared to its MSE's for $\beta_4$, $\beta_5$ and $\beta_6$. The simulation results seemed to be in agreement with the ones which we obtained from the MSCM data set applications.\\

\noindent [Insert Table 3 here.]\\ 

In the simulation study, we also included the traditional model in addition to the aforementioned two models. Related results (not shown here) seemed to be supporting the ones obtained while analyzing the MSCM data set, i.e., yielded same inferences with the common model. The R codes of the simulation study are available upon request from the authors. \\

\noindent \textbf{5 \ \ \ \ \ \ Discussion and conclusion} \\

In literature, it is common to consider separate covariate effects on multiple responses while constructing multivariate models. However, this assumption might be often redundant and too restrictive. In this article, we proposed a multivariate marginal modeling framework which permits building more flexible models. A user-friendly R package, {\bf mmm2}, was proposed to fit the model. Our modeling framework is not restricted to a specific response family, but handles several of them. Last but not least, all of the features of marginal model fitting with GEE are inherited for these multivariate models.

In this paper, we specifically considered first order GEE (GEE1) as proposed by ~\cite{liang86}, since our main focus was on the marginal mean parameters, i.e., the dependency parameters are of secondary interest. This version of GEE is known to yield inefficient estimates of the dependency parameters, since they are treated as nuisance parameters. If the scientific interest is on the dependency parameters together with the marginal mean parameters, second order GEE (GEE2 ~\cite{prentice88, liang92}) or alternating logistic regressions ~\cite{carey93} should be preferred. Throughout the paper, we considered GEE1 and called it GEE.\\

\newpage

\begin{table}[h]
\caption{List of the variables appear in the MSCM study and the related explanations.}
\label{tab:mscmvar}
\centering
\scalebox{0.90}{
\begin{tabular} {l l}
\hline
Variable & Explanation\\
\hline
id & id number of the mother and her child\\
stress & mother's stress status: 0=absence, 1=presence \\
illness & child's illness status: 0=absence, 1=presence \\
married & marriage status of the mother: 0=other, 1=married \\
education & mother's education level: 0=less than high school, \\
 & 1=high school graduate or more \\
employed & mother's employment status: 0=unemployed, 1=employed \\
chlth & child's health status at baseline: 0=very poor/poor, \\
 & 1=fair, 2=good, 3=very good \\
mhlth & mother's health status at baseline: 0=very poor/poor, \\
 & 1=fair, 2=good, 3=very good \\
race & child's race: 0=white, 1=non-white \\
csex & child's gender: 0=male, 1=female \\
housize & size of the household: 0=2-3 people, 1=more than 3 people\\
bstress & baseline stress: average value of the mother's stress \\
 & status for the first 16 days \\
billness & baseline illness: average value of the child's illness \\
 & status for the first 16 days \\
week & a time variable: calculated as (day-22)/7\\
\hline
\end{tabular}}
\end{table}

\newpage

\begin{table}[h]
\caption{Results of the MSCM data set application. Only robust standard error and Z estimates are reported. While Models 1, 2 were fitted under exchangeable working correlation structure, Model 3 was fitted under unstructured working correlation assumption.}
\label{tab:mscmresults1}
\centering
\scalebox{0.90}{
\begin{tabular} {l c c c}
\hline
 & Model 1  & Model 2 & Model 3 \\
\hline
 & Est. \ (SE) \ \ \ \ \ Z & Est. \ (SE) \ \ \ \ \ Z & Est. \ (SE) \ \ \ \ \ Z \\
\hline
Intercept       &  -2.14 \ (0.42) \ -5.15    & -2.23 \ (0.36) \ -6.12   & -2.58 \ (0.34) \ -7.49 \\
married         &  -0.01 \ (0.24) \ -0.02    & \ 0.25 \ (0.19) \ \ 1.34 & \ 0.22 \ (0.18) \ \ 1.19 \\
education       &   \ 0.36 \ (0.23) \ \ 1.62 & \ 0.19 \ (0.20) \ \ 0.94 & \ 0.25 \ (0.20) \ \ 1.27 \\
employed        &   -0.65 \ (0.25) \ -2.59   & -0.43 \ (0.22) \ -1.95   & -0.35 \ (0.22) \ -1.61 \\
chlth           &   -0.26 \ (0.13) \ -1.96   & -0.34 \ (0.12) \ -2.88   & -0.26 \ (0.11) \ -2.31 \\
mhlth           &   -0.17 \ (0.12) \ -1.39   & -0.11 \ (0.11) \ -0.97   & -0.18 \ (0.10) \ -1.73 \\
race            &   -0.02 \ (0.24) \ -0.06   & -0.01 \ (0.18) \ -0.04   & \ 0.19 \ (0.18) \ \ 1.02 \\ 
csex            &   -0.04 \ (0.22) \ -0.20   & \ 0.02 \ (0.18) \ \ 0.10 & \ 0.05 \ (0.17) \ \ 0.30 \\
housize         &   \ 0.06 \ (0.24) \ \ 0.26 & \ 0.04 \ (0.23) \ \ 0.15 & \ 0.17 \ (0.23) \ \ 0.74 \\
bstress         &   \ 3.89 \ (0.71) \ \ 5.48 & \ 3.48 \ (0.67) \ \ 5.22 & \ 3.59 \ (0.65) \ \ 5.53\\
billness        &   \ 0.86 \ (0.71) \ \ 1.21 & \ 1.52 \ (0.57) \ \ 2.65 & \ 1.51 \ (0.56) \ \ 2.68 \\
week            &  -0.43 \ (0.16) \ -2.65    & -0.31 \ (0.14) \ -2.20   & -0.36 \ (0.13) \ -2.72 \\
rtype           &   \ 0.56 \ (0.54) \ \ 1.04 & \ 0.89 \ (0.31) \ \ 2.91 & \ 1.03 \ (0.29) \ \ 3.55 \\
married*rtype   &   \ 0.50 \ (0.32) \ \ 1.57 &                          & \\
education*rtype &     -0.42 \ (0.31) \ -1.35  &                         &  \\
employed*rtype  &    \ 0.43 \ (0.38) \ \ 1.13 &                         &  \\
chlth*rtype     &    -0.14 \ (0.17) \ -0.82   &                         &  \\
mhlth*rtype     &    \ 0.20 \ (0.18) \ \ 1.12 &                         &  \\
race*rtype      &    \ 0.04 \ (0.32) \ \ 0.11 &                         &  \\
csex*rtype      &    \ 0.06 \ (0.29) \ \ 0.21 &                         &  \\
housize*rtype   &   -0.63 \ (0.32) \ -1.95   & -0.58 \ (0.30) \ -1.95   & -0.78 \ (0.29) \ -2.63 \\
bstress*rtype   &   -3.83 \ (1.10) \ -3.50   & -3.18 \ (0.99) \ -3.20   & -3.79 \ (0.95) \ -3.99 \\
billness*rtype  &   \ 1.32 \ (0.88) \ \ 1.50 & 							& \\
week*rtype      &   \ 0.24 \ (0.26) \ \ 0.91 &                          & \\
\hline
\end{tabular}}
\end{table}
 
\newpage

\begin{table}[h]
\caption{Results of the simulation study. Uns: unstructured, Exch: exchangeable, Ind: independence.}
\label{tab:simresults}
\centering
\scalebox{0.90}{
\begin{tabular} {c c c c c c c c c c c c}
\hline
\multicolumn{1}{c}{Parameter} & \multicolumn{1}{c}{True} & & \multicolumn{4}{c}{Parsimonious} & \multicolumn{4}{c}{Common} \\ \cline{2-9}
\hline
 & & & Uns & Exch & AR(1) & Ind & Uns & Exch & AR(1) & Ind \\
\hline
		  &         & \ Mean & \  \  -0.526  & -0.512  & -0.518  & -0.517  & \  \  -0.526  & -0.513  & -0.519  & -0.518 \\
$\beta_0$ & -0.500  & \ Bias & \  \  -0.026  & -0.012  & -0.018  & -0.017  & \  \  -0.026  & -0.013  & -0.019  & -0.018 \\
		  &  	    & \ MSE  & \  \ \ 0.005  & \ 0.004 & \ 0.004 & \ 0.004 & \  \  \ 0.008 & \ 0.008 & \ 0.008 & \ 0.008 \\ \hline
          &     	& \ Mean & \  \ \ 0.459  & \ 0.444 & \ 0.472 & \ 0.473 & \  \  \ 0.462 & \ 0.447 & \ 0.474 & \ 0.475 \\
$\beta_1$ & \ 0.500 & \ Bias & \  \ -0.041   & -0.056  & -0.028  & -0.027  & \  \  -0.038  & -0.053  & -0.026  & -0.025 \\
		  &  		& \ MSE  & \  \ \ 0.022  & \ 0.024 & \ 0.025 & \ 0.025 & \  \  \ 0.041 & \ 0.045 & \ 0.046 & \ 0.046 \\ \hline
		  &  		& \ Mean & \  \ \ 0.954  & \ 0.922 & \ 0.926 & \ 0.925 & \  \  \ 0.954 & \ 0.923 & \ 0.928 & \ 0.926 \\
$\beta_2$ & \ 0.900 & \ Bias & \  \  \ 0.054 & \ 0.022 & \ 0.026 & \ 0.025 & \  \  \ 0.054 & \ 0.023 & \ 0.028 & \ 0.026 \\ 
		  & 	    & \ MSE  & \  \  \ 0.011 & \ 0.008 & \ 0.009 & \ 0.009 & \  \  \ 0.018 & \ 0.016 & \ 0.016 & \ 0.016 \\ \hline
          &  		& \ Mean & \  \  \ 0.648 & \ 0.684 & \ 0.669 & \ 0.669 & \  \  \ 0.650 & \ 0.686 & \ 0.675 & \ 0.671 \\
$\beta_3$ & \ 0.600 & \ Bias & \  \ \ 0.048  & \ 0.084 & \ 0.069 & \ 0.069 & \  \  \ 0.050 & \ 0.086 & \ 0.075 & \ 0.071 \\
		  &         & \ MSE  & \  \ \ 0.046  & \ 0.051 & \ 0.053 & \ 0.054 & \  \  \ 0.086 & \ 0.096 & \ 0.100 & \ 0.099 \\ \hline
		  &  		& \ Mean & 		 & 		   & 		 & 		   & \  \  -0.002  &  -0.003 &  -0.001 &  -0.003 \\
$\beta_4$ & \ 0.000 & \ Bias & 		 & 		   & 		 & 	   	   & \  \  -0.002  &  -0.003 &  -0.001 &  -0.003 \\ 
		  &  		& \ MSE  & 		 & 		   & 		 & 		   & \  \  \ 0.014 & \ 0.014 & \ 0.014 & \ 0.014 \\ \hline
          &  		& \ Mean & 		 & 		   & 		 &		   & \  \  \ 0.000 & \ 0.001 & \ 0.001 & \ 0.001 \\
$\beta_5$ & \ 0.000 & \ Bias & 		 & 		   & 		 & 		   & \  \  \ 0.000 & \ 0.001 & \ 0.001 & \ 0.001 \\
		  &  		& \ MSE  & 		 & 		   & 		 &    	   & \  \  \ 0.073 & \ 0.084 & \ 0.086 & \ 0.085 \\ \hline
          & 	    & \ Mean & 		 & 	  	   & 		 &  	   & \  \  \ 0.002 & \ 0.003 & \ 0.001 & \ 0.003 \\
$\beta_6$ & \ 0.000 & \ Bias & 		 & 		   & 		 &  	   & \  \  \ 0.002 & \ 0.003 & \ 0.001 & \ 0.003 \\
		  & 	    & \ MSE  & 		 & 		   & 		 & 	   	   & \  \  \ 0.028 & \ 0.029 & \ 0.029 & \ 0.029 \\ \hline
		  &  		& \ Mean & 		 & 		   & 		 &  	   & \  \  \ 0.001 & \ 0.001 & -0.008  & \ 0.001 \\
$\beta_7$ & \ 0.000 & \ Bias & 		 & 		   & 		 & 		   & \  \  \ 0.001 & \ 0.001 & -0.008  & \ 0.001 \\
		  & 	    & \ MSE  & 		 & 		   & 		 & 		   & \  \  \ 0.157 & \ 0.177 & \ 0.180 & \ 0.177 \\
\hline
\end{tabular}}
\end{table}

\end{document}